\documentclass[10pt,final,journal,twocolumn]{IEEEtran}

\usepackage{verbatim}
\usepackage{todonotes}
\usepackage[cmex10]{amsmath}
\usepackage[compress]{cite}
\usepackage{graphicx}

\usepackage{amssymb}
\usepackage{multirow}
\usepackage{cite}
\usepackage{epsfig}
\usepackage{subcaption}

\usepackage{amsmath}
\usepackage{algorithm}

\usepackage{fancyhdr}
\usepackage{booktabs}
\usepackage{longtable,lscape}
\usepackage{eso-pic}

\usepackage[figuresright]{rotating} %
\usepackage{comment}	%
\usepackage{enumerate}	%
\usepackage{float}		%

\usepackage{threeparttable}

\usepackage{amssymb}
\usepackage{cite}
\usepackage{graphicx}

\graphicspath{{./fig/}}
\usepackage{bpchem}          %
\usepackage{textcomp}        %
\usepackage{multirow}        %
\usepackage{placeins}        %
\usepackage{array}           %
\usepackage{amsmath}

\usepackage{algpseudocode}

\usepackage[np]{numprint}
\npdecimalsign{\ensuremath{.}}   %

\npthousandsep{\,}       %
\npfourdigitnosep        %

\usepackage{color}

\begin{document}

\title{
Configurable Independent Component Analysis Preprocessing Accelerator
}

\author{Hsi-Hung Lu, Chung-An Shen, Mohammed E. Fouda, and Ahmed M. Eltawil

\thanks{Hsi-Hung Lu and Chung-An Shen are with Department of Electronic and Computer Engineering, National Taiwan University of Science and Technology,  Taipei, Taiwan. E-mail: cashen@mail.ntust.edu.tw.}
\thanks{Mohammed Fouda is with the Electrical Engineering and Computer Science Dept., University of California--Irvine, CA, USA. E-mail: foudam@uci.edu.}%
\thanks{Ahmed Eltawil is with King Abdullah University of Science and Technology, Thuwal, Saudia Arabia.}
}

\maketitle

\begin{abstract}
Independent component analysis (ICA) has been used in many applications, including self-interference cancellation for in-band full-duplex wireless systems and anomaly detection in industrial internet  of  things. This paper presents a high-throughput and highly efficient configurable preprocessing accelerator for the ICA algorithm. The proposed ICA accelerator has three major blocks that perform data centering, covariance matrix for computation, and eigenvalue decomposition (EVD). %
Specifically, the proposed accelerator is based on a high-performance matrix multiplication array (MMA). The proposed MMA architecture uses time-multiplexed processing so that the efficiency of hardware utilization is greatly enhanced. Furthermore, the processing flow utilizes parallel processing such that the centering, the calculation of the covariance matrix, and EVD are conducted simultaneously and are individually pipelined to maximize throughput. %
This paper presents the architecture, circuit design, and performance estimates based on post-layout extraction of the proposed preprocessing ICA accelerator. The proposed design achieves a throughput of 40.7~kMatrices per second at complexity of 73.3~kGE. %
\end{abstract}

\section{Introduction}
\label{ch1}

The blind source separation (BSS) problem lies in the separation of a mixed-signal which is a combination of many independent sources. For instance, the cocktail problem, where individual voice signals of multiple speakers need to be separated is a well-known BSS problem.  In the BSS problem, the mixed signal is constructed by mixing independent source signals through an unknown channel. Without prior knowledge of the mixing channel, independent source signals can be separated only by observing the mixed received signals \cite{Ref_Stone_2004}. In this context, independent component analysis (ICA) algorithm is one of the most widely used approaches for separating source signals in BSS problems \cite{Ref_Stone_2004,Ref_Yadaf_2015,Ref_Hyvarinen_1999}. The ICA algorithm separates the mixed-signal by taking advantage of the statistical independence and non-Gaussian properties of source signals \cite{Ref_Stone_2004}. The ICA algorithm has been applied in many applications. For example, it is proposed in \cite{Ref_Fouda_2020,Ref_Lu_2020} that the self-interference cancellation (SIC) for the In-Band Full-Duplex (IBFD) wireless communication system can be considered a BSS problem. The ICA algorithm is applied to separate the self-interference signal and the signal of interest. Furthermore, the ICA algorithm is applied to separate multiple superimposed signals to identify the cyclic features of each transmitted signal in cognitive radio (CR) system \cite{Ref_Mario_2018}. Recently, it is proposed in \cite{Ref_Huanzhuo_CDICA_2020, Ref_Huanzhuo_IEEENetwork_2021} that ICA can be used in a temporal-critical industrial internet of things (IoT) application for anomaly detection or identification. A common theme of these previous use cases is that the source signals are expected to be extracted in real-time thus demanding high-performance ICA architectures.

The ICA algorithm has two main stages: the preprocessing stage and the iterative stage \cite{Ref_Stone_2004,Ref_Hyvarinen_1999,Ref_Li_2010}. ICA preprocessing aims to whiten the received signal so that the correlation between the signals is reduced. The whitened received signal is sent to the iterative stage where the demixing matrix is estimated and the mixed signals are separated. The preprocessing stage is an essential step in ICA, as it enhances the quality of the separated signals and accelerates the convergence in the iterative stage. However, ICA preprocessing requires several highly complex vector and matrix computations, including the calculation of the covariance matrix and the eigenvalue decomposition (EVD). { According to our simulation profiling, discussed in detail in Section II.A, based on the system in \cite{Ref_Fouda_2020,Ref_Lu_2020}, preprocessing occupies between 17.3\% to 46.9\% of ICA processing time for different targeted SIC performance.} Therefore, designing an efficient preprocessing accelerator is a great challenge in realizing efficient ICA circuits and systems.

Several VLSI architecture designs for ICA preprocessors have been reported in the literature. Reference \cite{Ref_Van_2011} proposed an ICA-based signal separation architecture for a biomedical application. The preprocessor in this design is based on a serial processing flow, which results in a very low processing speed. { While in \cite{Ref_Yang_2015}, a low-power ICA processor was proposed to improve the processing throughput of the EVD in the ICA preprocessor based on a highly parallel systolic array structure. However, the area complexity of the preprocessor occupies about 90\% of the entire ICA system.} Furthermore, \cite{Ref_Van_2016} and \cite{Ref_Van_2019} reported an improved  ICA architecture that benefits from parallel processing. However, the processing flow of the proposed preprocessor is still inherently sequential in nature and thus, cannot achieve high throughput. %

A common drawback of many ICA preprocessors described in the literature is that they only support real-valued signals and operate at low processing speeds. However, new applications of ICA in use cases such as wireless communications \cite{Ref_Popovski_2018, Ref_Mukherjee_2018, Ref_Sutton_2019} where high data rates, latency-sensitive \cite{Ref_Huanzhuo_CDICA_2020, Ref_Huanzhuo_IEEENetwork_2021}, complex-valued samples are common demand novel, high throughput ICA architectures.  

This work presents a configurable, high-throughput, and highly efficient ICA preprocessing accelerator. The main contributions of this paper are summarized as follows:
\begin{itemize}
  \item A novel ICA preprocessing accelerator that can perform centering, covariance, and EVD is presented. The proposed architecture supports complex-valued signals whereas the number of received signals and sample lengths is configurable. %
  
  \item A high-performance matrix multiplication array (MMA) is presented. The proposed MMA architecture uses time-multiplexed processing, such that the hardware utilization is greatly enhanced. Therefore, compared to prior designs, the proposed MMA unit achieves high processing throughput with low area complexity.
  
  \item A low-latency centering unit and covariance unit based on the proposed MMA are presented. The operating flow is highly improved, so that the centering and calculating of the covariance matrix can be performed in parallel.
  
  \item A highly efficient EVD unit is proposed. The designed EVD engine is optimized for processing Hermitian matrices and has a low processing latency.
  
  \item Experimental results show that the proposed preprocessor outperforms prior designs. It achieves high throughput with reduced area and higher efficiency (defined as a throughput to gate count) compared to prior work.     
\end{itemize}

The remainder of this paper is organized as follows. The background of ICA and preprocessing for ICA are introduced in Section~\ref{ch2}. The proposed centering and covariance units are presented in Section~\ref{ch3}, and the proposed EVD unit is described in Section~\ref{ch4}. The experimental results and comparisons with prior designs are given in Section~\ref{ch5}. Conclusions are drawn in Section~\ref{ch6}.

\section{Background and Related Work}
\label{ch2}

This section introduces the basic concepts of the ICA. The state-of-the-art ICA designs with focus on preprocessors are also reviewed.

\subsection{The Basic Concept of ICA and the ICA Preprocessing}
Considering the BSS problem, the received signal is constructed by mixing independent source signals through an unknown channel. Without prior knowledge of the mixing channel, independent source signals can be separated only by observing the mixed received signals \cite{Ref_Stone_2004}. The ICA algorithm is the most widely used approach for separating source signals in BSS problems \cite{Ref_Stone_2004,Ref_Yadaf_2015,Ref_Hyvarinen_1999}. In the ICA algorithm, the mixed signals are separated by utilizing the statistical independence and non-Gaussianity of the source signals. Furthermore, the number of observed signals must be equal to the number of source signals, so in the generic ICA representation the mixed received signals and the source signals can be expressed in matrix form as follows:
\begin{equation}
	\mathbf{Y} = \begin{bmatrix} \mathbf{Y_1}\\\mathbf{Y_2} \\\vdots \\\mathbf{Y_N}\end{bmatrix}\!\! = \!\! 
		\begin{bmatrix}
			h_{11}	& h_{12} 		& \hdots   		& h_{1N}       	\\
			h_{21}		& h_{22}	& \hdots			& h_{2N}		  	\\
			\vdots & \vdots    		& \hdots  		& \vdots       	\\
			h_{N1} 		& h_{N2} 		& \hdots   		& h_{NN}  
		\end{bmatrix}\!\!
		\begin{bmatrix} \mathbf{X_{1}}\\\mathbf{X_{2}} \\\vdots \\\mathbf{X_{N}}\end{bmatrix} = \mathbf{HX} 
	\label{eq:ica_model}
\end{equation}
where $\mathbf{Y}$ and $\mathbf{X}$ are the mixed received signals and the source signals, respectively. { $\mathbf{Y_i}$ and $\mathbf{X_i}$ are both row vectors with $M$ elements where $M$ is the sample length.} The matrix $\mathbf{H}$ has size $N \times N$, where $N$ is the number of signals. After receiving the mixed received signals $\mathbf{Y}$, the ICA estimates a demixing matrix $\mathbf{W}$ and uses it to separate the received signals. The separated signals $\mathbf{\hat{X}}$ can be defined as follows:
\begin{equation}
\mathbf{\hat{X}} = { \begin{bmatrix} \mathbf{\hat{X_{1}}}\\\mathbf{\hat{X_{2}}} \\\vdots \\\mathbf{\hat{X_{N}}}\end{bmatrix} }\!\!=\!\!
		\begin{bmatrix}
			w_{11}	& w_{12} 		& \hdots   		& w_{1N}       	\\
			w_{21}		& w_{22}	& \hdots			& w_{2N}		  	\\
			\vdots & \vdots    		& \hdots  		& \vdots       	\\
			w_{N1} 		& w_{N2} 		& \hdots   		& w_{NN}  
		\end{bmatrix}\!\!
		{ \begin{bmatrix} \mathbf{Y_{1}}\\\mathbf{Y_{2}} \\\vdots \\\mathbf{Y_{N}}\end{bmatrix} } \!\!=\!\! \mathbf{WY} 
	\label{eq:ica_demix}
\end{equation}
where the demixing matrix $\mathbf{W}$ is the inverse of the mixing matrix, assuming the estimation is perfect.

In band full-duplex (IBFD) can be considered as a prime example application that demands high performance ICA architectures \cite{Ref_Fouda_2020,Ref_Lu_2020}, where the successive interference cancellation (SIC) in IBFD systems is considered as a BSS problem. In this case, the ICA algorithm is applied to separate the self-interference signal (SI) and the signal-of-interest (SOI). It is shown in \cite{Ref_Fouda_2020} that the received signal that contains the mixture of SOI and SI and a direct feedback digital transmitted signal are the input to the ICA algorithm. Thus, the BSS problem can be formulated as follows 
\begin{equation} \label{eq:C-FastICA_recv}
\begin{bmatrix} Y_{si} \\Y_{mix}\end{bmatrix}=
\begin{bmatrix} 1 & 0 \\H_{ord} & H_{soi} \end{bmatrix}
\begin{bmatrix} X_{si} \\X_{soi}\end{bmatrix}
\end{equation}
where $Y_{si}$ and $Y_{mix}$ are the two input signals to the ICA algorithm. The signal $Y_{si}$ is the direct input from the SI signal $X_{si}$ whereas $Y_{mix}$ is the mixture of the SI and the SOI signal $X_{soi}$. The $H_{ord}$ and $H_{soi}$ denote the SI and SOI channels respectively. The ICA algorithm then estimates the demixing matrix and separate the SI and SOI signals. The simulation results shown in \cite{Ref_Fouda_2020} illustrate that the output SINR of the ICA-based SIC scheme outperforms the conventional SIC approach. Furthermore, it is shown in \cite{Ref_Lu_2020} that different ICA algorithms such as FastICA \cite{Ref_Hyvarinen_1999} or ICA-EBM \cite{Ref_Li_2010} can be applied to this ICA-based SIC scheme.  

In ICA, the demixing matrix is estimated by searching for orthogonal vectors iteratively. Prior work in literature \cite{Ref_Stone_2004,Ref_Hyvarinen_1999,Ref_Li_2010} shows that including a preprocessing stage before searching for the orthogonal vectors is crucial for reducing the number of iterations required and for improving the quality of the estimated demixing matrix.
 
\begin{figure}[!t]
	\centering
	\includegraphics[width=0.19\textwidth]{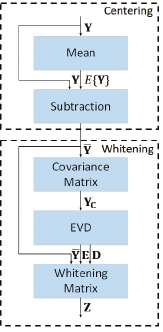}
	\caption{Operational flow in ICA preprocessing.}
	\label{fig_ICA_Preprocessing_Flow}
		\vspace{-0.1in}
\end{figure}

The main purpose of ICA preprocessing is to whiten the received signal as shown in Fig.~\ref{fig_ICA_Preprocessing_Flow}. In the first step of ICA preprocessing, the received signals are centered by subtracting the mean of signals, described as follows:
\begin{equation} \label{eq:Ybar}
\mathbf{\bar{Y}}_i = \mathbf{Y}_i - E\{ \mathbf{Y}_i \}
\end{equation}

Specifically, $\mathbf{\bar{Y}_i}$ and $E\{\mathbf{Y_i}\}$ are the $i$-th centered mixed received signal and its mean, respectively. The second step of preprocessing stage is whitening where the centered signals are whitened to have unit variance. In order to whiten the centered signals, the covariance matrix is calculated first as follows:
\begin{equation} \label{eq:Ycov}
\mathbf{Y_c} =  E\{ \mathbf{\bar{Y}\bar{Y}}^H \}
\end{equation}
\noindent where $\mathbf{Y_c}$ is the covariance matrix of the centered signals, which is also a $N \times N$ Hermitian matrix. In the following, the EVD is performed on the covariance matrix as follows:

\begin{equation} \label{eq:YcEVD}
\mathbf{Y_c} = \mathbf{ EDE }^H
\end{equation}

\noindent where $\mathbf{D}$ is a diagonal matrix composed of $N$ eigenvalues and $\mathbf{E}$ is an orthogonal matrix with $N$ eigenvectors. The matrices $\mathbf{D}$ and $\mathbf{E}$ are applied to whiten $\mathbf{\bar{Y}}$ as follows. 

\begin{equation} \label{eq:whitening}
\mathbf{Z} = \mathbf{ D^{-1/2} E^H \bar{Y} }
\end{equation}


{ The normalized execution time has been profiled based on the experimental platform in Matlab for the SIC in IBFD system \cite{Ref_Fouda_2020,Ref_Lu_2020}. This iteration part of the SIC is based on the FastICA algorithm. The normalized execution time for each process is tabulated in Table~\ref{Tab:Timing_ratio} where the number in this table represents the ratio in percentile over total processing time (including preprocessing and iteration). It can be seen from Table~\ref{Tab:Timing_ratio} that for the platform shown in \cite{Ref_Fouda_2020,Ref_Lu_2020}, the average number of iterations is approximately equal to 25 to achieve a post-SIC SINR of 22dB. Thus, the preprocessing time occupies about 17.3\% of the entire SIC execution time. In other words, the preprocessing engine needs to preprocess the frame within 17.3\% of the frame duration time. Considering the system in \cite{Ref_Fouda_2020,Ref_Lu_2020} has a 8.19~ms frame duration time over 512 symbols and 48 subcarriers, the preprocessing time should be less than 29.5~\textmu s. Thus, it becomes clear that as the number of iterations increases the amount of time available for preprocessing reduces and the need for an ultra high throughput preprocessor becomes essential. On the other hand, when the number of iterations is reduced, the percentage of time needed for preprocessing dominates the performance, further stressing the need for high throughput preprocessing.} 

\begin{table}
\centering
\caption{ {The Normalized Execution Time (\%) for FastICA Algorithm}}
\begin{tabular}{|c|c|c|c|}
\hline
     {{Processes}} & {Iteration=5} & {Iteration=15} & {Iteration=25} \\
\hline
     {{Centering}} & {11.3} & {6.2} & {4.2} \\
\hline
     {{Covariance}} & {11.1} & {5.9} & {4.1}  \\
\hline
     {{EVD}} &  {9.1} &  {4.8} &  {3.3} \\
\hline
     {{Whitening}} & {15.4} & {8.4} & {5.7} \\
\hline
     {{Preprocessing}} & {46.9} & {25.3} & {17.3} \\
\hline
     {{Iteration}} & {53.1} & {74.7} & {82.7} \\
\hline
\end{tabular} 

\label{Tab:Timing_ratio}
\vspace{-0.1in}
\end{table}

\subsection{Related Work}

Numerous VLSI architectures for ICA and ICA preprocessors have been reported in the literature. Reference \cite{Ref_Van_2011} proposes a real-valued signal separation architecture based on the FastICA algorithm. The preprocessor in this architecture has a centering unit, a covariance unit, and an EVD engine. Since the units in this preprocessor operate in a very sequential manner, the latency is extended. Specifically, the centering unit and covariance unit sequentially access the same memory several times, leading to very long delays. According to our analysis, for  $N$ complex-valued  received signals and a sample length of $M$, a total of $8 \times N \times M + 6 \times M \times N \times N $ cycles are consumed by the centering and covariance units of \cite{Ref_Van_2011}. Furthermore, to increase the throughput, parallelism is introduced in the preprocessor in \cite{Ref_Chen_2020}. An MMA based on a systolic array structure is proposed in this design for calculating the covariance matrix. This MMA uses a real-valued matrix with a size of $3 \times 64$. It takes three cycles to multiply a column vector and a row vector. However, the utilization of the MMA is not optimized, especially for complex-valued matrices. In addition, in \cite{Ref_Van_2016} and \cite{Ref_Van_2019}, the preprocessor has to share computing units, resulting in excessive processing delays.

Moreover, the EVD engine is one of the core units in the ICA preprocessor and has a high degree of complexity. An EVD unit in the preprocessor based on the approximate Jacobi algorithm was presented in \cite{Ref_Yang_2015}. This EVD unit has very long critical path and results in low operating frequency. A configurable SVD unit based on the CORDIC operation and folded systolic array structure was described in \cite{Ref_Chen_2020_SVD}. Although these authors produced a low-complexity circuit, the processing had high latency and was slow. Furthermore, a sorting network was proposed in \cite{Ref_Shahshahani_2020} to enhance the efficiency of memory accesses in the EVD architecture. However, a matrix multiplier was employed in this design to perform matrix rotations, which resulted in high area complexity.

Modern wireless communication systems operate at a data rate of hundreds of megabits per second \cite{Ref_Popovski_2018, Ref_Mukherjee_2018, Ref_Sutton_2019} with complex-valued signals. Furthermore, IoT applications usually require strict real-time processing \cite{Ref_Huanzhuo_CDICA_2020, Ref_Huanzhuo_IEEENetwork_2021}. Therefore, it is crucial to design a high-throughput ICA preprocessor for these systems. No relevant improvements for the operational flow and architecture of ICA preprocessors have been reported in the literature. To achieve a high-throughput and highly efficient ICA preprocessor, in this work, we design the architecture by jointly considering the operational flow and circuit structure for the centering, covariance, and EVD units. The proposed centering and covariance units are operated in parallel so that the processing throughput is much higher. Furthermore, the covariance unit is based on a proposed novel highly efficient MMA circuit. To reduce the amount of computation, a novel diagonalization process for a Hermitian matrix was designed for the EVD. In addition, we would like to point out that the proposed preprocessor can be used in different ICA algorithms that have been proposed in literature such as \cite{Ref_Stone_2004,Ref_Hyvarinen_1999,Ref_Li_2010}.

{ 
\section{The Proposed Centering and Covariance Units}
\label{ch3}

\subsection{The Proposed Efficient MMA Architecture}\label{sec:3a}

\begin{figure}[!t]%
	\centering
	\begin{tabular}{c}
	\includegraphics[width=0.85\columnwidth]{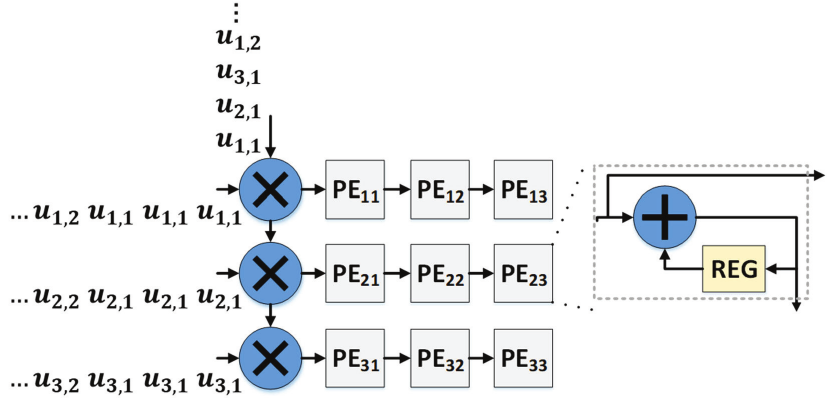} \\
	(a) \\ 
	\includegraphics[width=0.9\columnwidth]{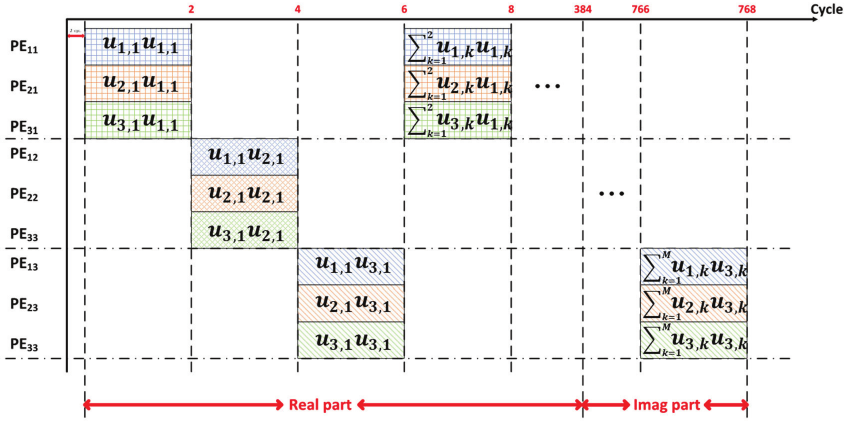}  \\
	(b)  \\
	\end{tabular}
	\caption{((a) The architecture of the Matrix Multiplication Unit in \cite{Ref_Chen_2020} and (b) the operational flows of applying this MMA for complex-valued matrix }
	\label{fig_MMA_Original}
		\vspace{-0.1in}
\end{figure}

\begin{figure}[!t]%
	\centering
	\begin{tabular}{c}
	\includegraphics[width=0.85\columnwidth]{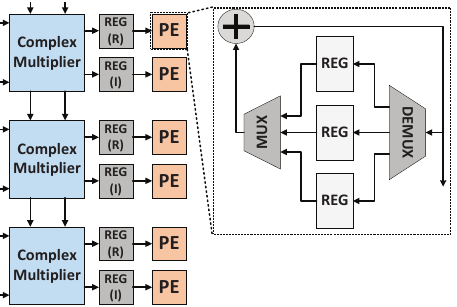} \\
	(a) \\ 
	\includegraphics[width=0.85\columnwidth]{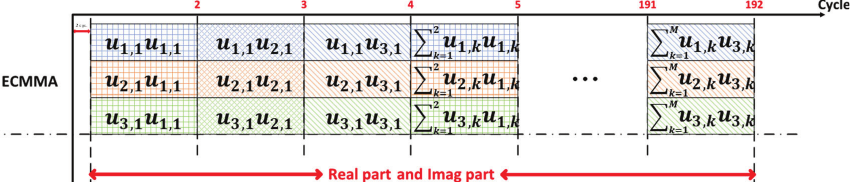}  \\
	(b)  \\
	\end{tabular}
	\caption{(a) Architecture of the proposed ECMMA and (b) operational flows of the conventional and proposed MMAs.}
	\label{fig_MMA_Circuit}
		\vspace{-0.1in}
\end{figure}

\begin{table}[!tb]
\centering
\caption{Comparison of MMAs}\label{tab:comp}
\begin{tabular}{lcc}
\hline
      & {MMA of \cite{Ref_Chen_2020} } & {Proposed MMA} \\
     \hline
     {Matrix size} & $m \times n$ & $m \times n$ \\
     {PEs} & $2 \times m^{2}$ & $2 \times m$  \\
     {Adders} & $2 \times m^{2}$ & $2 \times m$ \\
     {Multiplexors} & 0 & $4 \times m$  \\
     {Registers} & $2 \times m^{2} + 2 \times m$ & $2 \times m^{2} + 2 \times m$ \\
     {Cycles} & $m \times n$ & $m \times n$\\
     \hline
\end{tabular}
\label{Tab:MMA_Comparison}
\end{table}

{ A matrix multiplication unit (MMA) is presented in \cite{Ref_Chen_2020} to compute the multiplication of a real-valued $3\times64$ matrix with its transpose. The structure of this MMA is shown in Fig.~\ref{fig_MMA_Original}(a) for calculating equations as follows

\small
\begin{equation} \label{eq:MMA}
\begin{gathered}
\begin{bmatrix}
u_{1,1} & \cdots & u_{1,64} \\
u_{2,1} & \cdots & u_{2,64} \\
u_{3,1} & \cdots & u_{3,64} \\
\end{bmatrix} \times
\begin{bmatrix}
u_{1,1} & u_{2,1} & u_{3,1} \\
\vdots & \vdots & \vdots \\
u_{1,64} & u_{2,64} & u_{3,64} \\
\end{bmatrix} \\
= \begin{bmatrix}
u_{1,1}u_{1,1} + \cdots + u_{1,64}u_{1,64} & \cdots & u_{1,1}u_{3,1} + \cdots + u_{1,64}u_{3,64} \\ 
u_{2,1}u_{1,1} + \cdots + u_{2,64}u_{1,64} & \cdots & u_{2,1}u_{3,1} + \cdots + u_{2,64}u_{3,64} \\ 
u_{3,1}u_{1,1} + \cdots + u_{3,64}u_{1,64} & \cdots & u_{3,1}u_{3,1} + \cdots + u_{3,64}u_{3,64} \\ 
\end{bmatrix}
\end{gathered}
\end{equation}
\normalsize

The operation flow of applying this MMA \cite{Ref_Chen_2020} to a complex-valued $3\times64$ matrix is shown in Fig.~\ref{fig_MMA_Original}(b). It is shown in Fig.~\ref{fig_MMA_Original}(b) that the PEs are not efficiently utilized as they are idled in multiple cycles. Furthermore, there is only one multiplier in each row as shown in Fig.~\ref{fig_MMA_Original}(a) and four real-valued multiplications are needed for one complex-valued multiplication. As a result, a total of $3 \times 64 \times 4 = 768$ cycles are consumed.

In this work, an efficient complex-valued MMA (ECMMA) is employed in centering and covariance units in the preprocessor where the architecture is shown in Fig.~\ref{fig_MMA_Circuit}(a) assuming a $3\times64$ matrix. It is noted that the architecture, shown in Fig.~\ref{fig_MMA_Circuit}(a), is based on the example of a $3\times64$ matrix for a fair comparison with the design shown in \cite{Ref_Chen_2020}. It is shown in Fig.~\ref{fig_MMA_Circuit}(a) that three complex-valued multipliers are used instead of the original real-valued multiplier to process the real and imaginary parts of the covariance matrix. Since the proposed ECMMA directly performs complex-valued operations, the real and imaginary parts of the elements in the covariance matrix are computed concurrently. Thus, one cycle is consumed by the PE in the accumulation of the real and imaginary parts of the element. Furthermore, it is shown in Fig.~\ref{fig_MMA_Circuit}(a) that one column of PEs is designed where three registers and one multiplexer and demultiplexer are contained in the PE. Specifically, each PE in the proposed MMA is utilized to compute and store the multiply-and-accumulate (MAC) for different columns of data using time-multiplexing. Figure~\ref{fig_MMA_Circuit}(b) shows operational flows for the proposed ECMMA assuming the $3\times64$ matrix. It can be seen from Fig.~\ref{fig_MMA_Circuit}(b) that due to the time-multiplexed utilization of the PE, the efficiency of the proposed MMA is much higher. In particular, a total of $3 \times 64 = 192$ cycles are consumed.}

Table~\ref{tab:comp} compares the number of hardware elements in the proposed efficient MMA and the MMA in \cite{Ref_Chen_2020} for an $m\times n$ complex-valued matrix. This table shows that, although additional multiplexers are needed, the number of adders is much lower. Moreover, the number of adders required for the MMA of \cite{Ref_Chen_2020} increases exponentially with the size of the matrix, whereas the number of adders for the proposed efficient MMA increases linearly.

\subsection{The Proposed Centering and Covariance Units}

As mentioned in Section~\ref{sec:3a}, the prior centering and covariance units sequentially access the memory multiple times and incur excessive delays. To achieve high-throughput preprocessing for the ICA, this work optimizes the operational flow so that the centering unit and the covariance unit run in parallel. Furthermore, the covariance unit is based on the proposed efficient ECMMA, which strikes a balance between high speed and low complexity. In particular, $N$ received signals lead to a $N\times N$ Hermitian covariance matrix, and the proposed operational flow is based on the characteristic of the Hermitian matrix. The main idea is that the Hermitian covariance matrix has three different parts, and the elements in each part are generated at different stages by different approaches. Figure~\ref{fig_Covariance_Matrix_Example} is an example of an $8\times8$ Hermitian covariance matrix $\mathbf{Y_c}$ assuming eight received signals where $\mathbf{\bar{Y}_i}$ denotes the $\mathbf{i}$-th row vector in the centered received signals $\mathbf{\bar{Y}}$.   
\begin{figure}
	\centering
	\includegraphics[width=0.5\textwidth]{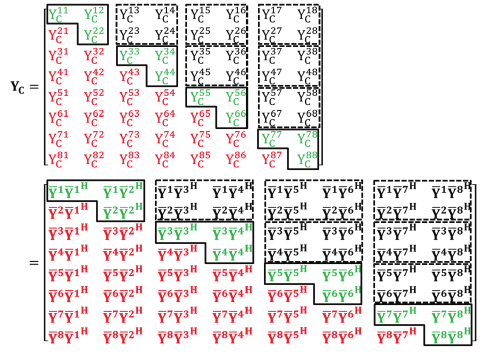}
	\caption{Example of a covariance matrix with eight received signals.}
	\label{fig_Covariance_Matrix_Example}
		\vspace{-0.15in}
\end{figure}
The lower-triangular elements highlighted in red in Fig.~\ref{fig_Covariance_Matrix_Example} do not need to be calculated. Furthermore, the elements highlighted with solid lines, such as ${Y_{c11}}$, ${Y_{c12}}$, and ${Y_{c22}}$, can be computed once the centered signals $\mathbf{\bar{Y}_1}$ and $\mathbf{\bar{Y}_2}$ are generated. Therefore, in this design, a novel processing flow is proposed so that the elements highlighted with solid lines are computed concurrently with the operation of the centering unit while other centered signals, such as $\mathbf{\bar{Y}_3}$ and $\mathbf{\bar{Y}_4}$, are being generated. Finally, the remaining upper-triangular elements, such as those highlighted with dotted lines, are decomposed as $2 \times 2$ submatrices and each submatrix is processed independently. { Considering a case with $N$ received signals, the number of $2 \times 2$ submatrices is $(N^2 - 2 \times N)/8$. The proposed design is based on the processing flow with a high degree of regularity and can be applied to any covariance Hermitian matrix with an even number of received signals.} 

Based on the proposed processing flow, the architecture of the centering and covariance units is presented in Fig.~\ref{fig_Architecture_Centering_Covariance}. 
\begin{figure*}[!t]\hspace*{-2.2em}
	\centering
	\includegraphics[width=14cm, height=4.5cm]{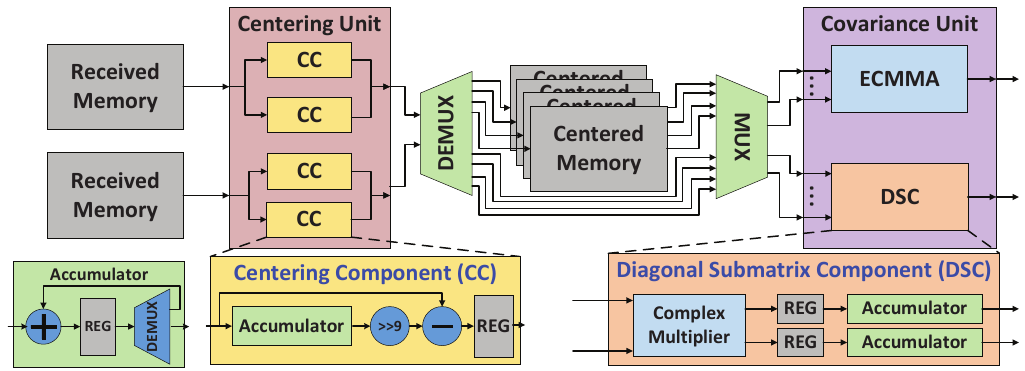} \\
	\caption{Architecture of the proposed centering unit and covariance unit.}
	\label{fig_Architecture_Centering_Covariance}
		\vspace{-0.1in}
\end{figure*}
\begin{figure*}[!hbt]\hspace*{-2.2em}
	\centering
	\includegraphics[width=18cm, height=4cm]{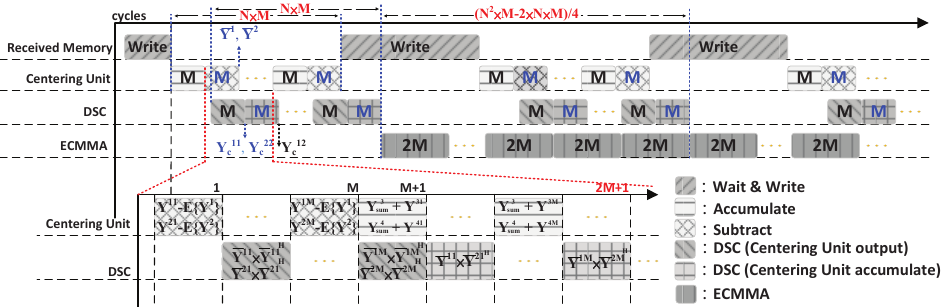} \\
	\caption{Timing diagram of the proposed centering and covariance flow.}
	\label{fig_Centering_Covariance_Flow}
		\vspace{-0.15in}
\end{figure*}
It is shown in Fig.~\ref{fig_Architecture_Centering_Covariance} that, for a $2\times2$ matrix, the centering unit has four centering components (CCs) and the covariance unit has one diagonal submatrix component (DSC) and one ECMMA. The DSC computes those elements shown in solid lines in Fig.~\ref{fig_Covariance_Matrix_Example}, whereas the ECMMA unit computes the elements shown in dashed lines. Furthermore, the four CCs compute two complex-valued signals concurrently. Each CC sequentially accumulates data from the input signal and calculates the average value. The same signal is then read by the next CC in the sequence and the average value is subtracted to give the centered signal. The centered signal is output in sequence and is both written into the memory and sent to the DSC in the covariance unit. Therefore, based on the proposed processing flow, the centering unit outputs the centered data in sequence while the DSC in the covariance unit calculates elements of the covariance matrix at the same time. After all the received signals have been centered and the elements in each L-shaped region marked by a solid line in Fig.~\ref{fig_Covariance_Matrix_Example} have been calculated by the DSC, the elements in the squares marked with a dotted line continue to be calculated through the ECMMA in the covariance unit. { Given $N$ received signals, the ECMMA unit calculates the $(N^2 - 2 \times N)/8$ submatrices in sequence.} Furthermore, it can be seen that the centering and covariance operations overlap in time and the elements of covariance matrix are generated as soon as possible. Therefore, the throughput of the overall architecture is greatly enhanced.

Figure~\ref{fig_Centering_Covariance_Flow} shows a detailed timing diagram for the proposed centering and covariance units assuming $N$ received signals and the sample length is $M$. After taking $(N\times M)/2$ cycles to write received signals into the received memory, row vectors $\mathbf{{Y}_1}$ and $\mathbf{{Y}_2}$ are read by the four CCs in the centering unit in sequence to calculate the average values where $M$ cycles are consumed. The vectors $\mathbf{{Y}_1}$ and $\mathbf{{Y}_2}$ are read again and the average values are subtracted to generate the centered signals of $\mathbf{{\bar{Y}}_1}$ and $\mathbf{{\bar{Y}}_2}$ by taking $M$ cycles. { It is shown in Fig.~\ref{fig_Centering_Covariance_Flow} that $N \times M$ cycles are consumed by the centering unit given $N$ received signals and the sample length is $M$.} { The centered signals are saved into the memory and sent to the DSC of the covariance unit at the same time. While the centering unit is centering the data, the DSC takes $M$ cycles to calculate ${Y}_{c11}$ and ${Y}_{c22}$ for the covariance matrix. In particular, the $N \times N$ Hermitian covariance matrix is decomposed as diagonal submatrices and 2x2 off-diagonal submatrices, whereas the DSC computes diagonal terms in the covariance matrix. The block diagram of DSC is shown in Fig.~\ref{fig_Architecture_Centering_Covariance}, which is composed of a complex multiplier and two accumulators. Since the diagonal elements is the multiplication of the variable with its complex conjugate, the result must be a real number. Thus, the imaginary part of the computation can be omitted. Therefore, the computation of ${Y}_{c11}$ and ${Y}_{c22}$ can be calculated concurrently.} Subsequently, the centering unit reads vectors $\mathbf{{Y}_3}$ and $\mathbf{{Y}_4}$ sequentially and performs the accumulation. Since the centering unit does not write any data to the memory at this time, the DSC reads $\bar{Y}_{r1}$, $\bar{Y}_{r2}$, $\bar{Y}_{i1}$, and $\bar{Y}_{i2}$ from the memory and calculates ${Y}_{c12}$ for the covariance matrix. It can be observed from Fig.~\ref{fig_Centering_Covariance_Flow} that while the centering operations are being performed on all $N$ complex signals, $(N/2) \times 3$ out of $((1+N) \times N)/2$ elements in the covariance matrix (i.e., those in the L-shaped region marked by a solid line in Fig.~\ref{fig_Covariance_Matrix_Example}) are computed at the same time.  { A total of $N \times M$ cycles are consumed by the DSC given $N$ received signals and the sample length is $M$.}  

The remaining covariance elements are decomposed into $(N^2-2\times N)/8$) $2\times2$ submatrices and are processed by the ECMMA unit. It is shown in Fig.~\ref{fig_Centering_Covariance_Flow} that $2\times M$ cycles are consumed by the ECMMA unit for calculating four elements concurrently. To maximize the throughput, the centering unit and the DSC can start to process the next $N \times M$ complex matrix while the ECMMA unit is processing. Figure~\ref{fig_Centering_Covariance_Flow} shows that the processing of a $2 \times 2$ ECMMA for the first matrix and the processing of the centering and covariance units overlap for the second matrix. {The covariance matrix is generated every $(N^2\times M-2 \times N \times M)/4$ cycles assuming $N \geq 8$. For example, the covariance matrix is computed every 6144 cycles for an $8 \times 512$ complex-valued matrix.} {Therefore, based on the architecture and the operational flow, a balance between the high-throughput and low-complexity can be achieved.}  

\begin{figure*}
	\centering
	\includegraphics[width=1.2\columnwidth]{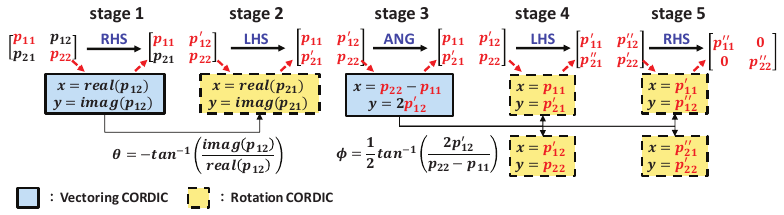} \\
	\caption{Proposed processing sequence for a $2 \times 2$ complex-valued Hermitian submatrix EVD diagonalization.}
	\label{fig_EVD_2by2}
	\vspace{-0.15in}
\end{figure*}

\begin{figure*}
	\centering
	\includegraphics[width=0.7\textwidth, height=5.5cm]{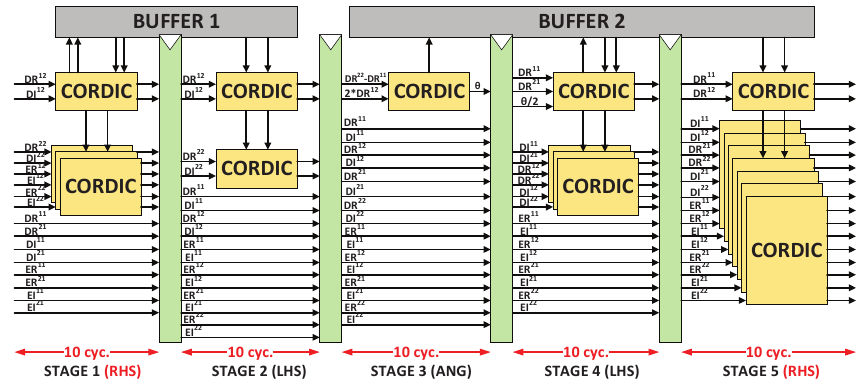} \\
	\caption{Architecture of the proposed EVD engine.}
	\label{fig_EVD_PE}
	\vspace{-0.15in}
\end{figure*}

{
\subsection{Configurability of Centering and Covariance Units }
It is noted that the proposed processing flow of the centering and covariance units shown in Fig.~\ref{fig_Centering_Covariance_Flow} is applied to any covariance Hermitian matrix with an even number of received signals without changing the architecture presented in Fig. ~\ref{fig_Architecture_Centering_Covariance}. In other words, the same architecture is applied with different matrices sizes whereas the operating cycles are adaptive to the size of signals. To be specific, given $N$ received signals and the sample length is $M$, the proposed centering unit is operated to accumulate $M$ samples and to calculate the mean. The centered result is computed by subtracting the mean value for each sample. Thus, it is shown in Fig.~\ref{fig_Centering_Covariance_Flow} that $N \times M$ cycles are consumed by the centering unit presented in Fig. ~\ref{fig_Architecture_Centering_Covariance} to calculate centered results. Similarly, $N \times M$ cycles are consumed by the DSC in Fig. ~\ref{fig_Architecture_Centering_Covariance} to calculate the diagonal values. Furthermore, considering $N$ received signals, a $N \times N$ Hermitian covariance matrix has resulted. The number of $2 \times 2$ off-diagonal submatrices is then $(N^2 - 2 \times N)/8$ according to Fig.~\ref{fig_Covariance_Matrix_Example}. The ECMMA unit presented in Fig. ~\ref{fig_Architecture_Centering_Covariance} calculates the $(N^2 - 2 \times N)/8$ submatrices in sequence. Thus, based on the architecture shown in Fig. ~\ref{fig_Architecture_Centering_Covariance}, the covariance matrix is generated every $(N^2\times M-2 \times N \times M)/4$ cycles assuming $N \geq 8$. Therefore, to process with different sizes of input data, the controller and finite state machine (FSM) in the proposed design are realized such that data sizes are treated as parameters and the operating cycles are adaptive to those parameters. For example, the covariance matrix is computed in every 6144 cycles for an $8 \times 512$ complex-valued matrix and in every 5120 cycles for an $10 \times 256$ complex-valued matrix. 
}

\section{Proposed EVD Unit}
\label{ch4}
\subsection{Proposed EVD Design}
The most commonly used approach for EVD is based on the Jacobi algorithm \cite{Ref_Shahshahani_2020,Ref_Yang_2015} where a series of rotation transformations is exploited to find the eigenvalues and eigenvectors of a matrix. Specifically, in the Jacobi algorithm, an $N \times N$ matrix is decomposed into multiple $2 \times 2$ submatrices and the submatrices that are on the diagonal of the matrix (known as diagonal submatrices) are diagonalized. Those non-diagonal submatrices are rotated according to the angles of the diagonalization process. This diagonalization-rotation process is performed iteratively.

\begin{figure*}
     \centering
     \begin{subfigure}[b]{0.3\textwidth}
         \centering
         \includegraphics[width=\textwidth]{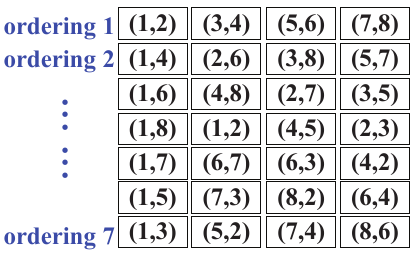}
         \caption{}
     \end{subfigure}
     \hfill
     \begin{subfigure}[b]{0.3\textwidth}
         \centering
         \includegraphics[width=1.2\textwidth]{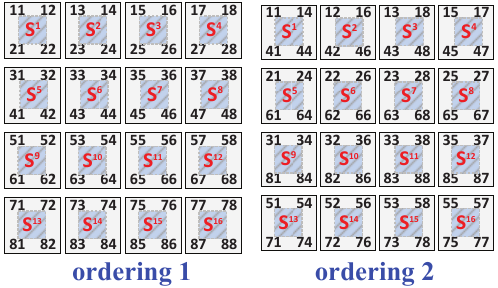}
         \caption{}
     \end{subfigure}
     \hfill
     \begin{subfigure}[b]{0.3\textwidth}
         \centering
         \includegraphics[width=0.65\textwidth]{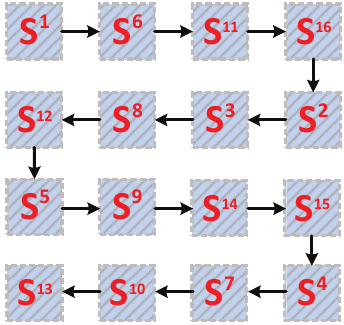}
         \caption{}
     \end{subfigure}
        \caption{(a) Parallel ordering \cite{Ref_Chen_2020_SVD}. (b) The matrix reconstructed based on the parallel ordering. (c) The proposed processing sequence for each ordering.}
        \label{fig_Parallel_Ordering}
        \vspace{-0.15in}
\end{figure*}

In the ICA preprocessing, the input to the EVD unit is the Hermitian covariance matrix. The covariance matrix is decomposed into multiple $2 \times 2$ submatrices and, to achieve an optimized balance between throughput and complexity, the proposed EVD unit fully utilizes the Hermitian properties in the design of the processing sequence and the architecture for the diagonalization and rotation of a $2 \times 2$ submatrix. The proposed processing sequence is presented in Fig.~\ref{fig_EVD_2by2}. It requires five stages for the EVD diagonalization of a $2 \times 2$ submatrix. Note that this $2 \times 2$ submatrix is the diagonal submatrix of the covariance matrix, so it must be Hermitian. In other words, the non-diagonal elements $p_{12}$ and $p_{21}$ in Fig.~\ref{fig_EVD_2by2} are complex conjugates of each other. The processing sequence is based on calculating the angle and the matrix rotation. Stages 1 and 2 eliminate the imaginary parts of $p_{12}$ and $p_{21}$ in the submatrix. The rotation angle $\theta$ is computed accordingly. In the following stage 3, the corresponding rotation angle $\phi$ is computed. Finally, stages~4 and 5 eliminate the non-diagonal elements of the submatrix, resulting in a diagonal $2\times2$ submatrix. Note that, since the $2 \times 2$ submatrix is a Hermitian matrix, the diagonal elements have real values. As a result, all the elements in the submatrix will be real after stage 2. Furthermore, since the non-diagonal elements are complex conjugates of each other, the rotation angle $\theta$ needs to be computed only once.

Figure~\ref{fig_EVD_PE} presents the architecture of the proposed EVD unit to process the flow shown in Fig.~\ref{fig_EVD_2by2}. There are CORDIC elements in each stage, and it is architected as a 10-cycle pipeline. In this figure, DR and DI are the real and imaginary parts of the eigenvalues, and ER and EI are the real and imaginary parts of the eigenvectors, respectively. A subscript represents the corresponding position in the submatrices. $D$ is initialized to the value of the submatrix, and $E$ is initialized to the identity matrix. The diagonal $2 \times 2$ submatrices are sent into the EVD unit one by one. In stage 1, the first CORDIC operates in vectoring mode and obtains the right-hand side (RHS) rotation direction. The corresponding RHS rotations are performed by the other three CORDICs based on the direction provided by the first CORDIC. Thus, the imaginary part of $D_{12}$ is eliminated. Since $D_{12}$ and $D_{21}$ are complex conjugates of each other, in stage 2, two CORDICs perform the left-hand side (LHS) rotation on the submatrix using the same rotation direction as stage 1. The imaginary part of $D_{21}$ is eliminated in stage 2 and the diagonal $2 \times 2$ submatrix is transformed into a real symmetric matrix. In stage 3, the CORDIC calculates the rotation angle to eliminate the non-diagonal elements of the submatrix for the subsequent stages. In stage 4, the first CORDIC performs the LHS rotation with the rotation angle calculated in stage 3 and provides the rotation direction for the other three CORDICs. In stage 5, eight CORDICs perform the RHS rotation to eliminate the non-diagonal elements based on the rotation direction generated by stage 4. Therefore, each diagonal $2 \times 2$ submatrix can be converted into a real-valued diagonal matrix. After all diagonal $2 \times 2$ submatrices have been so converted, the non-diagonal $2 \times 2$ submatrices are input one by one, and the corresponding rotation is performed by the EVD unit operating in rotation mode. The rotation of a non-diagonal $2 \times 2$ submatrix is through the corresponding rotation direction and angle.

\subsection{EVD Processing Sequence}
{The EVD of an $N\times N$ matrix is an iterative process involving the diagonalization of $2\times 2$ diagonal submatrices and the corresponding rotations of $2\times 2$ non-diagonal submatrices. How those $2\times 2$ diagonal and non-diagonal submatrices are constructed and the processing order affect the efficiency of the EVD \cite{Ref_Shahshahani_2020}. To accelerate the iterative process, a concept of parallel ordering \cite{Ref_Chen_2020_SVD} has been widely applied and is illustrated in Fig.~\ref{fig_Parallel_Ordering}(a) for the example of an $8 \times 8$ matrix.

According to the parallel ordering \cite{Ref_Chen_2020_SVD}, the $8 \times 8$ matrix is decomposed into seven orderings, each of which has four $2\times 2$ diagonal submatrices. These submatrices can be diagonalized simultaneously without inferfering with each other. The notation $(i,j)$ in Fig.~\ref{fig_Parallel_Ordering}(a) denotes the $2 \times 2$ diagonal submatrix with elements at $(i,i)$, $(i,j)$, $(j,i)$, and $(j,j)$ in the $8 \times 8$ matrix. For example, given the $8\times8$ matrix with elements $h_{ij}$ where $i$ is the row and $j$ the column, the notation $(1,2)$ represents the $2\times2$ submatrix of elements $[h_{11}, h_{12} ; h_{21}, h_{22}]$. In each ordering, the diagonalization processes for the $2\times 2$ diagonal submatrices are independent of each other, and thus, can be conducted in parallel. For example. the diagonalization of the four $2 \times 2$ submatrices $(1,2)$, $(3,4)$, $(5,6)$, $(7,8)$ are conducted in parallel in ordering 1.   

In this design, each ordering is processed by the proposed EVD engine in a pipelined fashion. However, the sequence for rotating the non-diagonal submatrices also affects the efficiency of the EVD and needs to be optimized. To reduce the latency and improve the throughput, a novel processing flow for the EVD is also proposed in this work. Figure~\ref{fig_EVD_PE} shows that the proposed EVD unit has five stages so that five $2\times2$ submatrices are processed concurrently in a pipeline. To achieve this, the ordering for processing different submatrices needs to be specially designed. In particular, each ordering is processed based on the result from the prior ordering. For example, the calculations for ordering 2 cannot start until all the submatrices of ordering 1 have been output. This results in processing delays between the orderings. To reduce the latency, this design proposes a processing sequence such that the data dependency between two orderings can be eliminated and the efficiency can be significantly enhanced. Consider an $8\times8$ matrix. The matrix that is reconstructed based on the parallel ordering is illustrated in Fig.~\ref{fig_Parallel_Ordering}(b). Orderings~1 and~2 are shown as examples. Each $2\times2$ submatrix in the reconstructed matrix of Fig.~\ref{fig_Parallel_Ordering}(b) is denoted as $S_{i}$. The proposed processing sequence for processing the $2\times2$ submatrix in  each ordering is shown in Fig.~\ref{fig_Parallel_Ordering}(c).} 

{ Figure~\ref{fig_EVD_PE_Timing} depicts the timing analysis diagram for EVD for an $8 \times 8$ matrix using the proposed EVD architecture and based on the proposed sequence shown in Fig.~\ref{fig_Parallel_Ordering}. The submatrix is input to the EVD unit according to the parallel ordering and the designed processing sequence for every ten cycles. The main purpose of the proposed processing sequence is to make the beginning of the processing for each new ordering and the end of the processing for each previous ordering independent. Figure~\ref{fig_EVD_PE_Timing} shows that the processing of the last four submatrices for ordering 1 and the processing of the first four submatrices for ordering 2 can be performed in parallel. This greatly enhances the efficiency of the proposed pipelined EVD engine. Before the calculation of ordering 2, there must be a 40-cycle idle time to avoid a data hazard. However, by sorting the input submatrices, ordering 2 can be calculated before the submatrices of ordering 1 are completely output and no data hazard will occur.} Finally, the last block to perform the whitening is presented in Fig.~\ref{fig_WU}. The CORIC unit in this block calculates the inverse square root of the diagonal entries of the D matrix and the $2 \times 2$ ECMMA units compute the multiplications of the matrices.

\begin{figure}
	\centering
	\includegraphics[width=\columnwidth]{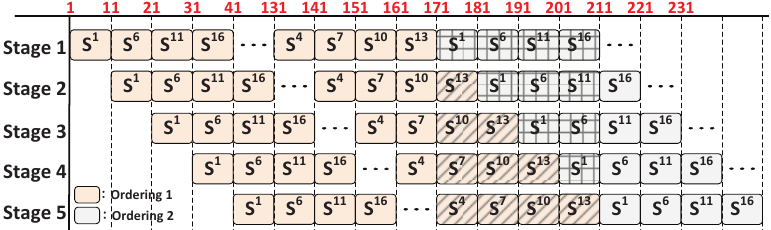} \\
	\caption{Timing analysis diagram for the EVD of an $8\times8$ matrix using the proposed EVD engine.}
	\label{fig_EVD_PE_Timing}
\end{figure}

\begin{figure}
	\centering
	\includegraphics[width=\columnwidth]{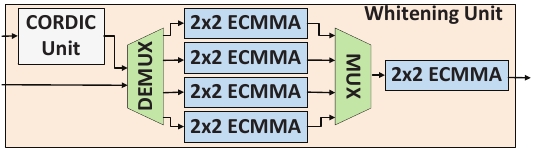} \\
	\caption{The architecture of the whitening unit.}
	\label{fig_WU}
\end{figure}
}
{
\subsection{Configurability of EVD and Whitening Units }
The proposed EVD operational flow and timing analysis shown in Fig.~\ref{fig_Parallel_Ordering} and Fig.~\ref{fig_EVD_PE_Timing} can support different matrices sizes without changing the hardware structure. In other words, the controller and FSM of the proposed EVD unit control the operating cycles according to the parameter of $N$. For an $N \times N$ matrix, the $N/4$ $2 \times 2$ submatrices are sent into the EVD unit to be processing according to the order shown in Fig.~\ref{fig_Parallel_Ordering}(c). Considering an $N \times N$ matrix, the total processing time for the EVD unit can be approximately calculated as $((N^2/4) \times (N-1)) \times 10$. Similarly, the whitening unit processes with different parameters of $N$ using different numbers of clock cycles based on the same hardware structure. Once the EVD Unit is completed, the CORDIC in the whitening unit calculates the value of the diagonal item in the D matrix in $N \times 32$ cycles. Finally, the four $2 \times 2$ ECMMA units in the whitening unit compute the multiplications of the matrices by approximately $(N/8) \times N \times M$ cycles.    
}

\section{Experimental Results and Comparisons}
\label{ch5}

\subsection{Implementation Results}

\begin{figure}
	\centering
	\includegraphics[width=\columnwidth]{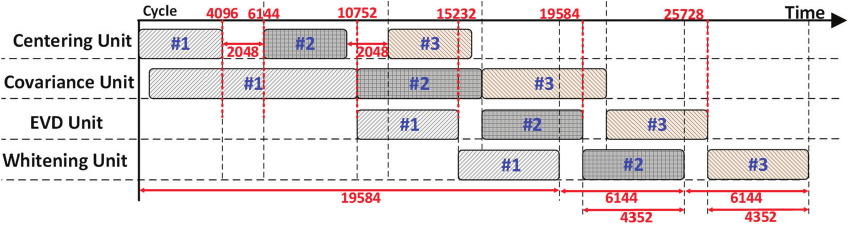} \\
	\caption{Timing schedule of the proposed preprocessor.}
	\label{fig_Total_Timing}
		\vspace{-0.15in}
\end{figure}

\begin{figure}
	\centering
	\includegraphics[width=0.75\columnwidth]{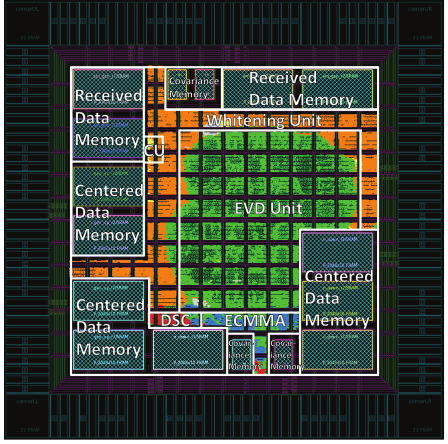}
	\caption{Layout of the proposed preprocessor.}
	\label{fig_layout}
	\vspace{-0.1in}
\end{figure}

Based on the proposed operational flow and architecture, an ICA preprocessor was designed and implemented.  The preprocessor was synthesized, and placed-and-routed using the TSMC 90~nm process with a nominal power supply voltage of 1.1~V. { Table~\ref{Tab:Implementation} reports estimates for the operating frequency, latency, throughput, and complexity based on the post-layout simulations. It is noted that the proposed architecture is configurable for different numbers of signals $N$ and sample lengths $M$. The memory size limits the largest size of the sample lengths $M$ that can be supported and the estimates summarized in Table~\ref{Tab:Implementation} is based on the memory size supports up to a sample length of 512. Thus, the area complexity of this preprocessor is 73.3~kGE in terms of two-input NAND gates with 30~kB of on-chip single-port SRAM. Figure~\ref{fig_Total_Timing} is the complete timing diagram for the implemented preprocessor assuming number of signals is equal to eight and the sample length is 512. The figure shows that 6144 and 4480 cycles are consumed for the centering and covariance units and the EVD unit, respectively. Furthermore, the whitening unit is operated in parallel with the other units. Therefore, a throughput of 40.7~kMatrices per second with a latency of 78.33~\textmu{}s is achieved given the operating frequency of 250~MHz.} { The processing time can be scaled with the values of $N$ and $M$ according to the timing diagram of Figs~\ref{fig_Centering_Covariance_Flow},~\ref{fig_EVD_PE_Timing}, and ~\ref{fig_Total_Timing}. For example, assuming $N$ = 16 and $M$ = 512, the latency will be 94976 cycles for the proposed preprocessor. The corresponding latency and throughput are summarized in Table~\ref{Tab:Implementation}.} { Figure ~\ref{fig_layout} shows the layout of the proposed preprocessor including centering, covariance, EVD, and whitening units of which architectures are illustrated in Fig. ~\ref{fig_Architecture_Centering_Covariance}, Fig. ~\ref{fig_EVD_PE}, and Fig. ~\ref{fig_WU}.}

\begin{table}
\centering
\caption{Implementation results for the proposed preprocessor with different sizes of input signals and sample lengths.}
\begin{tabular}{lc}
\hline
Parameter & Value \\
\hline
     {Technology} & TSMC 90 nm \\
     {Frequency} & 250 MHz  \\
     {Word length} & 10-bit fixed point  \\
     {Core area} & \np{810928}~\textmu m$^{2}$ \\
     {Gate count}$^\text{a}$ & 73.3 kGE \\
     {Memory} & 30 kB single-port SRAM \\
     {Matrix size} & { $4 \times 256$ / $4 \times 512$ / $8 \times 256$ / $8 \times 512$} \\
     {Latency (cycles)} &  { \np{3424} / \np{6240}  / \np{12160}  /  \np{19584} }  \\
     {Latency (\textmu s)} &  {  13.69 /  24.96 /  48.64 /  78.33 }  \\
     {Throughput (kMatrices/s)} & {244.1 / 121.1 / 55.8 / 40.7}\\
     \hline
\end{tabular} 

   $^\text{a}$Gate count is core area / NAND gate area. \\
\label{Tab:Implementation}
\vspace{-0.1in}
\end{table}

\begin{table*}[!t]
\centering
\caption{Comparisons of the centering and covariance units for different designs of preprocessor}
\begin{tabular}{lccccccc}
    \hline
    & \multicolumn{2}{c}{{\cite{Ref_Van_2011} }} & \multicolumn{2}{c}{{\cite{Ref_Yang_2015}}} & \multicolumn{2}{c}{{\cite{Ref_Van_2019}} } & {This}  \\
    {Parameter} & Original & Scaled & Original & Scaled & Original & Scaled & work \\
    \hline
    {Technology (nm)} & 90 & 90 & 90 & 90 & 90 & 90 & 90\\
    {Architecture type} & Real & Real & Real & Real & Real & Real & Complex\\
    {Arithmetic type} & Fixed & Fixed & Fixed & Fixed & Floating & Floating & Fixed\\
    {Frequency (MHz)} & 100 & 100 & 11 & 11 & 100 & 100 & 250 \\
    {Matrix type} & Real & Complex & Real & Complex & Real & Complex & Complex\\
    {Matrix size} & $8 \times 256$ & $8 \times 512$ & $8 \times 256$ & $8 \times 512$ & $8 \times 1024$ & $8 \times 512$ & { $4 \times 256$ / $4 \times 512$ / $8 \times 256$ / $8 \times 512$} \\
    {Latency (cycles)} 
    & 24576$^\text{a}$ & 196608$^\text{b}$ & 24576 & 196608$^\text{b}$ & 24576$^\text{a}$ & 24576$^\text{b}$ & { 1792 / 3584 / 5376 / 10752} \\   %
    {Throughput (\textmu{}Matrices/cycle)} & 40.7$^\text{a}$  & 5.1$^\text{b}$ & 40.7$^\text{a}$ & 5.1$^\text{b}$ & 40.7$^\text{a}$ & 40.7$^\text{b}$ & { 976.56 / 488.28 / 223.21 / 162.76} \\
    {Throughput (kMatrices/s)} & 4.07$^\text{a}$ & 0.51$^\text{b}$ & 0.45$^\text{a}$ & 0.06$^\text{b}$ & 4.07$^\text{a}$ & 4.07$^\text{b}$ & { 244.1 / 122.1 / 55.8 / 40.7} \\
    
    {Complexity (kGE)} & 2.76$^\text{a}$ & 2.76$^\text{a}$ & 1.16$^\text{a}$ & 1.16$^\text{a}$ & 486$^\text{a}$ & 486$^\text{a}$ & 11.8 \\

    {Power (mW)} &  16.35$^\text{c}$ & n/a & 0.0816$^\text{c}$ & n/a & 65.0$^\text{c}$ & n/a & 50.2 \\
    {Efficiency}$^\text{d}$ & 14.75 & 1.85 & 35.086 & 4.397 & 0.08 & 0.08 & { 82.759 / 41.380 / 18.916 / 13.793} \\ %
    \hline
\end{tabular} \\[1ex]
   $^\text{a}$Latency and complexity are analyzed proportionally according to the timing and architecture diagram reported in the literature. The latency and complexity for the design \cite{Ref_Van_2011} is analyzed based on Fig.~2 and and~18 in \cite{Ref_Van_2011}. Furthermore, the throughput and complexity for \cite{Ref_Yang_2015} are analyzed based on Table~I in \cite{Ref_Yang_2015}. The latency and complexity for the design \cite{Ref_Van_2019} is analyzed based on Fig.~3.2 and Fig.~4.5 in \cite{Ref_Van_2019}. \\
   $^\text{b}$Latency and throughput are scaled based on the processing of the complex-valued $8\times512$ matrix. It is assumed that the architecture remains the same and the processing time is scaled for processing larger complex-valued matrices. The memory is scaled for storing a complex-valued $8\times512$  matrix. \\
   $^\text{c}$Power consumption is for the entire chip, not just the centering and covariance units. \\
   $^\text{d}$Efficiency = Throughput (\textmu{}Matrices/cycle) / Complexity (kGE).
\label{Tab:Cent_Comp}
\end{table*}

\begin{table*}[!h]
\centering
\caption{Comparison of EVD units for different designs of preprocessor}
\scalebox{0.9}{
\begin{tabular}{lcccccccccccc}
\hline
     & \multicolumn{2}{c}{{\cite{Ref_Van_2011, Ref_Van_2019}} }  & \multicolumn{2}{c}{{\cite{Ref_Yang_2015}} }  & \multicolumn{2}{c}{{\cite{Ref_Chen_2020_SVD}} }  & \multicolumn{2}{c}{{\cite{Ref_Shahshahani_2020}} }  & \multicolumn{2}{c}{{\cite{Ref_Sajjad_2021}} }  & This \\
     {Parameter} & Original & Scaled & Original & Scaled & Original & Scaled & Original & Scaled & Original & Scaled & work \\
     \hline
     {Technology (nm)} & 90 & 90 & 90 & 90 & 90 & 90 & 90 & 90 & - & - & 90 \\
     {Architecture type} & Real & Real & Real & Real & Complex & Complex & Real & Real & Real & Real & Complex\\
     {Arithmetic type} & Floating & Floating & Fixed & Fixed & Fixed & Fixed & Fixed & Fixed & Fixed & Fixed & Fixed\\
     {Frequency (MHz) } & 100 & 100 & 11 & 11 & 500 & 500  & 813 & 813 & 10 & 10 & 250  \\
     {Matrix type} & Real & Complex & Real & Complex & Complex & Complex & Real & Complex & Real & Complex & Complex\\
     {Matrix size} & $8\times8$ & $8\times8$ & $8\times8$ & $8\times8$ & $8\times8$ & $8\times8$ & $8\times8$ & $8\times8$ & $8\times8$ & $8\times8$ &{$4\times4$ / $8\times8$} \\
     
     {Latency (cycles)}               & 
      112896$^\text{a}$  &  225792$^\text{b}$  & 
      1680              &   3360$^\text{b}$   &  
      1336$^\text{a}$    &   1336$^\text{a}$  &  
      700$^\text{a}$     &  1400$^\text{b}$    & 
      840               &  1680$^\text{b}$    & 
      { 480 / 4480} \\
      {SA throughput$^\text{c,e}$  } 
     & 8.9$^\text{a}$ & 4.43$^\text{b}$ 
     & 595.24$^\text{a}$ & 297.62$^\text{b}$ 
     & 2200 & 2200 
     & 8929.89 & 4464.95$^\text{b}$ 
     & 1190$^\text{a}$ & 595$^\text{b}$ 
     & {2083.333 / 223.214}  \\
     {SA throughput$^\text{c,f}$  } 
     & 0.89$^\text{a}$ & 0.443$^\text{b}$ 
     & 6.55$^\text{a}$ & 3.27$^\text{b}$ 
     & 1100 & 1100 
     & 7620 & 3630$^\text{b}$ 
     & 11.9$^\text{a}$ & 5.95$^\text{b}$ 
     & {520.8 / 55.8}  \\
     
     {ICA throughput$^\text{c,e}$  } 
     & n/a & 4.43$^\text{b}$ 
     & n/a & 162.76$^\text{b}$ 
     & n/a & 162.76$^\text{b}$ 
     & n/a & 162.76$^\text{b}$ 
     & n/a & 162.76$^\text{b}$ 
     & {976.56 / 488.28 / 223.21 / 162.76}  \\
     {ICA throughput$^\text{c,f}$ } 
     & n/a & 0.443$^\text{b}$ 
     & n/a & 1.79$^\text{b}$ 
     & n/a & 81.38$^\text{b}$ 
     & n/a & 132.32$^\text{b}$ 
     & n/a & 1.6276$^\text{b}$ 
     & {244.1 / 122.1 / 55.8 / 40.7}  \\
     
     {Complexity (kGE) } & 76.5$^\text{a}$ & 76.5$^\text{a}$ & 61.2$^\text{a}$ & 61.2$^\text{a}$ & 192 & 192  & 481 & 481 & 156$^\text{a}$ & 156$^\text{a}$ & 61.5 \\
     Power (mW) & 16.35$^\text{d}$ & n/a & 0.0816$^\text{d}$ & n/a & 112$^\text{d}$ & n/a  & n/a & n/a & 152$^\text{d}$ & n/a & 50.2 \\

     {SA efficiency} %
     & 0.12 & 0.06
     & 9.726 & 4.863 
     & 11.458 & 11.458 
     & 18.565 & 9.283 
     & 7.6 & 3.8 
     & {33.875 / 3.629}    \\

     {ICA efficiency} %
     & n/a & 0.06 
     & n/a & 2.659 
     & n/a & 0.848 
     & n/a & 0.338 
     & n/a & 1.04 
     & {15.879 / 7.940 / 3.629 / 2.647}   \\
     \hline
     
\end{tabular}
}  \\[1ex]

   $^\text{a}$The latency for \cite{Ref_Van_2011} and \cite{Ref_Van_2019} are analyzed based on Fig.~4 in \cite{Ref_Van_2011}, and the complexity is analyzed based on Fig.~4.5 in \cite{Ref_Van_2019}. Furthermore, the latency and complexity of \cite{Ref_Yang_2015} and \cite{Ref_Chen_2020_SVD} are analyzed based on Table~I in \cite{Ref_Yang_2015} and Table~I and Figs.~5 and~6 in \cite{Ref_Chen_2020_SVD} respectively. The latency for \cite{Ref_Shahshahani_2020} is analyzed based on the calculations mentioned in Section~IV-B in \cite{Ref_Shahshahani_2020} and the throughput of \cite{Ref_Sajjad_2021} is computed based on Table~3 of \cite{Ref_Sajjad_2021}. \\
   $^\text{b}$The latency and throughput are scaled based on the EVD of a complex-valued $8\times8$ matrix. %
   \\
     $^\text{c}$Standalone (SA) is where the throughput is estimated by assuming only the EVD is operating. ICA is where the throughput is calculated by assuming the centering and covariance units reported in this work are the front stages of the EVD. \\
     $^\text{d}$The power is for the entire chip. $^\text{e}$The unit is \textmu{}Matrices/cycle. $^\text{f}$The unit is kMatrices/s. \\
\label{Tab:EVD_Comp}
\end{table*}

\begin{table*}[!t]
\centering
\caption{Comparison with state-of-the-art ICA preprocessors}
\begin{tabular}{lccccccc}
\hline
      &
      \multicolumn{2}{c}{{\cite{Ref_Van_2011} }} & 
      \multicolumn{2}{c}{{\cite{Ref_Yang_2015}}} & 
      \multicolumn{2}{c}{{\cite{Ref_Van_2019}}} & 
      {This} \\
     {Parameter} & Original & Scaled & Original & Scaled & Original & Scaled &  work \\
    \hline

     {Technology (nm)} & 90 & 90 & 90 & 90 & 90 & 90 & 90\\
     {Architecture type} & Real & Real & Real & Real & Real & Real & Complex\\
     {Arithmetic type} & Hybrid & Hybrid & Fixed & Fixed & Floating & Floating & Fixed\\
     {Frequency (MHz)} & 100 & 100 & 11 & 11 & 100 & 100 & 250 \\
     {Matrix type} & Real & Complex & Real & Complex & Real & Complex & Complex\\
     {Matrix size} & $8\times256$ & $8\times512$ & $8\times256$ & $8\times512$ & $8\times1024$ & $8\times512$ & {$4\times 256$ / $4\times512$ / $8\times 256$ / $8\times512$}\\
     
    {Latency (cycles) }
    &  137472$^\text{a}$  &  422400$^\text{a}$   
    &  26256$^\text{a}$  &  199968$^\text{a}$  
    &  137472$^\text{a}$  &  250368$^\text{a}$  
    &  {\np{3424} / \np{6240} / \np{12160} / \np{19584}}  \\
    
    {Throughput (\textmu{}Matrices/cycle) } 
    & 7.27$^\text{a}$ & 2.37$^\text{a}$ 
    & 38.09$^\text{a}$ & 5.001$^\text{a}$ 
    & 7.27$^\text{a}$ & 3.99$^\text{a}$ 
    & {976.56 / 488.28 / 223.21 / 162.76} \\
    
    {Throughput (kMatrices/s) } 
    & 0.727$^\text{a}$ & 0.237$^\text{a}$ 
    & 0.419$^\text{a}$ & 0.055$^\text{a}$ 
    & 0.727$^\text{a}$ & 0.399$^\text{a}$ 
    & {244.1 / 122.1 / 55.8 / 40.7} \\

    {Memory (kB) } &  8$^\text{a}$  & 32$^\text{a}$ & 3.5$^\text{a}$  & 14$^\text{a}$ & 32$^\text{a}$ & 32$^\text{a}$ & 30 \\
    
    {Logic complexity (kGE) } 
    & 79.26$^\text{a}$ & 79.26$^\text{a}$ 
    & 62.36$^\text{a}$ & 62.36$^\text{a}$ 
    & 562.5$^\text{a}$ & 562.5$^\text{a}$ 
    & 113.3 \\
    
    {Memory complexity (kGE) } 
    & 50.81$^\text{c}$ & 129.91$^\text{c}$ 
    & 24.7$^\text{c}$ & 63.75$^\text{c}$ 
    & 129.91$^\text{c}$ & 129.91$^\text{c}$ 
    & 123.51 \\
    
    {Total complexity (kGE) } 
    & 130.07$^\text{c}$ & 209.17$^\text{c}$ 
    & 87.06$^\text{c}$ & 126.11$^\text{c}$ 
    & 692.41$^\text{c}$ & 692.41$^\text{c}$ 
    & 236.81 \\
    
    {Power (mW) } &  16.35$^\text{b}$ & n/a & 0.0816$^\text{b}$ & n/a & 65.0$^\text{b}$ & n/a & 58.2  \\

    Efficiency 
    & 0.06 & 0.011
    & 0.437 & 0.04 
    & 0.01 & 0.006 
    & {4.124 / 2.062 / 0.943 / 0.687} \\
    \hline

\end{tabular}  \\[1ex]

   $^\text{a}$Latency, complexity, and memory are analyzed based on Tables~\ref{Tab:Cent_Comp} and~\ref{Tab:EVD_Comp}. Throughput is calculated based on the processing latency. \\
   $^\text{b}$Power is for the entire chip, not just the preprocessor. \\
   $^\text{c}$The memory block is generated, and the complexity is estimated based on the area of two-input NAND gates. The total complexity is the sum of the logic complexity and the memory complexity. \\  
\label{Tab:Total_Comp}
\vspace{-0.05in}
\end{table*}

\subsection{Comparison with Prior Designs}

Table~\ref{Tab:Cent_Comp} compares the proposed centering and covariance units with prior work \cite{Ref_Van_2011,Ref_Yang_2015,Ref_Van_2019}. Since the implementation results reported in the literature are for the entire ICA system, the latency and complexity for the centering and covariance parts of the system were estimated for the prior designs. For example, the latency of the centering and covariance unit for the design in \cite{Ref_Van_2011} was estimated according to Fig.~2 in that paper. Furthermore, according to Figs.~2 and~18 in \cite{Ref_Van_2011}, the area complexity was estimated to be 2.76~kGE. The centering and covariance units reported in \cite{Ref_Yang_2015} are similar to those in \cite{Ref_Van_2011}, and so the latency is the same. However, the area complexity was estimated according to Tables~I and~IV in \cite{Ref_Yang_2015} due to a difference in the data widths. Moreover, the design of \cite{Ref_Van_2019} processes real-valued $8\times1024$ data using four parallel computing units. The complexity is based on Table~I and the layout of Fig.~4.5 in \cite{Ref_Van_2019}. In addition, since the designs in \cite{Ref_Van_2011}, \cite{Ref_Yang_2015}, and \cite{Ref_Van_2019} use serialized processing, the throughput was calculated as the processing latency.

{ For a fair comparison, the latencies for the prior designs were scaled for a complex-valued $8\times512$ matrix.} We assumed that the architectures of the prior designs were unchanged. The processing time was scaled according to the type and size of the matrix. A real-valued multiplier executes four times for a complex-valued multiplication. Thus, since the matrix size is doubled and the matrix type is complex-valued, the latencies were increased by eight times compared to the designs of \cite{Ref_Van_2011} and \cite{Ref_Yang_2015}. Furthermore, since the design in \cite{Ref_Van_2019} has multipliers operating in parallel, the latency was the same for a reduced-size complex matrix. 

Table~\ref{Tab:Cent_Comp} shows that the proposed design achieves much higher throughput and lower latency compared with the designs of \cite{Ref_Van_2011} and \cite{Ref_Yang_2015}, although at the cost of slightly increased area complexity. This is because these designs are based on serialized processing architectures, whereas the proposed covariance and centering units operate in a parallel and pipelined fashion. Moreover, the floating-point design of \cite{Ref_Van_2019} has a faster processing speed because it has four parallel computing units. While the accuracy is enhanced by the floating-point operation, the processing delay and area complexity are also greatly increased. 

To illustrate the balance between throughput and complexity, efficiency was used as the figure of merit (FOM) in the comparisons \cite{Ref_Chen_2020_SVD,Ref_Shahshahani_2020}. Efficiency is defined as the ratio of throughput and complexity indicating the performance gain for the cost. Table~\ref{Tab:Cent_Comp} shows that, due to the highly optimized processing flow and the highly efficient utilization of hardware resources, the proposed design achieves the best efficiency compared to the prior designs. 

Table~\ref{Tab:EVD_Comp} compares the proposed EVD architecture with prior work. The latency for the EVD in the prior work was estimated if the results reported in the literature are based on the entire ICA system. In particular, the EVD in \cite{Ref_Van_2011} and \cite{Ref_Van_2019} is based on a single floating-point CORDIC component, so the latency and complexity are analyzed according to Fig.~4 in \cite{Ref_Van_2011} and Fig.~4.5 in \cite{Ref_Van_2019}. Furthermore, the complexity of the design of \cite{Ref_Yang_2015} was analyzed according to Table~I in \cite{Ref_Yang_2015}. On the other hand, the SVD unit in \cite{Ref_Chen_2020_SVD} has a fully pipelined architecture, so the latency was analyzed based on Fig.~6 in \cite{Ref_Chen_2020_SVD}. The EVD unit proposed in \cite{Ref_Shahshahani_2020} is based on a structure containing four PEs, so the latency was calculated according to Section~IV-B in that work. Moreover, the latency for processing a real-valued $8\times8$ matrix in the design of \cite{Ref_Sajjad_2021} is reported in Table~3 in \cite{Ref_Sajjad_2021}. The complexity was estimated based on Tables~2 and~5 in that work.

{ For a fair comparison, the latency for the prior designs was scaled for a complex-valued $8\times8$ matrix, as shown in Table~\ref{Tab:EVD_Comp}.} We again assumed that the architectures were unchanged. The processing time was scaled according to the size and the type of the matrix. For example, the latency of \cite{Ref_Van_2011} was increased by two times for the EVD of the complex-valued $8\times8$ matrix since the Jacobi algorithm operates on the real and imaginary parts separately. Furthermore, to evaluate the performance of EVD in ICA preprocessing, the throughput was also estimated by assuming that the centering and covariance units proposed in this work are the front stages of the EVD unit, which is referred to as ICA throughput in Table~\ref{Tab:EVD_Comp}. The throughput for \cite{Ref_Van_2011} was calculated by assuming that the EVD operates by itself, which is referred to as SA throughput. Since the proposed centering and covariance units output a result every 6144 cycles, the throughput is still dominated by the latency of EVD, so ICA throughput was the same as SA throughput for that design. The latency for \cite{Ref_Yang_2015} was doubled for the complex-valued matrix, and SA throughput was calculated accordingly. Since the architecture of \cite{Ref_Chen_2020_SVD} was designed for a complex-valued matrix, the scaled latency is the same as the original latency. The scaled latency was doubled and the scaled SA throughput was halved for the designs in \cite{Ref_Shahshahani_2020} and \cite{Ref_Sajjad_2021} because of the complex-valued EVD. However, the scaled ICA throughput for \cite{Ref_Yang_2015, Ref_Shahshahani_2020, Ref_Sajjad_2021} was limited by the input of the matrix from the centering and covariance units.

It can be seen from Table~\ref{Tab:EVD_Comp} that compared to \cite{Ref_Van_2011}, the proposed EVD unit achieves a higher throughput with reduced complexity due to the highly optimized pipelined processing flow. Furthermore, the architecture in \cite{Ref_Yang_2015} is based on the systolic array structure, so its latency is greatly reduced. While the latency of \cite{Ref_Yang_2015} in terms of clock cycles is lower than the proposed design, the throughput is limited by the very low operating frequency. Moreover, the fully pipelined design in\cite{Ref_Chen_2020_SVD} achieves high SA throughput because multiple matrices can be processed at the same time. However, the throughput is degraded for ICA preprocessing. On the other hand, the proposed EVD unit is specifically optimized for ICA, and compared with \cite{Ref_Chen_2020_SVD}, it strikes the best balance between throughput and complexity. In addition, the architecture of \cite{Ref_Shahshahani_2020} maximizes the throughput by using multiple fully pipelined CORDIC elements and matrix multipliers. Compared to \cite{Ref_Shahshahani_2020}, the throughput of the proposed EVD unit is lower while the complexity is also much reduced. The design in\cite{Ref_Sajjad_2021} leads to a heavily degraded throughput due to the low operating frequency. The efficiency FOM is also used to compare the proposed EVD in ICA preprocessing by the different designs. Table~\ref{Tab:EVD_Comp} shows that the proposed design outperforms the other designs in terms of efficiency. This is attributed to the highly optimized processing flow and the highly efficient utilization of hardware resources.

Table~\ref{Tab:Total_Comp} compares the proposed preprocessor with the prior ICA preprocessor designs. { For a fair comparison, the latencies for the prior designs were scaled for a complex-valued $8\times512$ matrix.} The latency and complexity of the ICA are due to the centering and covariance units as well as the EVD unit, as shown in Tables~\ref{Tab:Cent_Comp} and~\ref{Tab:EVD_Comp}. Furthermore, the total complexity in Table~\ref{Tab:Total_Comp} includes the logic as well as the memory. Specifically, to store an $8\times512$ complex-valued matrix, the size of the memory in \cite{Ref_Van_2011} and \cite{Ref_Yang_2015} has to be increased by four times. Since the design in \cite{Ref_Van_2019} has multipliers operating in parallel, the size of the memory is the same for a reduced-size complex matrix. Table~\ref{Tab:Total_Comp} shows that the preprocessor of \cite{Ref_Van_2011} results in low throughput due to the serialized processing flow.  Furthermore, the EVD in \cite{Ref_Van_2011} has a floating-point structure, which results in high complexity. Moreover, the throughput of \cite{Ref_Yang_2015} is degraded due to the lower efficiency of the processing flow and the low operating frequency. In addition, due to its fully floating-point design, the throughput of \cite{Ref_Van_2019} is higher at the cost of greatly increased area complexity. Table~\ref{Tab:Total_Comp} shows that the proposed design outperforms the other designs in terms of the efficiency FOM. This is attributed to the highly optimized processing flow and the highly efficient utilization of hardware resources. It is noted that the proposed design has a higher power consumption compared to the designs of \cite{Ref_Van_2011} and \cite{Ref_Yang_2015} due to the higher operating frequency and the support of complex-valued operations. Furthermore, the fully floating-point design \cite{Ref_Van_2019} leads to a high power consumption albeit a lower operating frequency. 

{
\subsection{Case Study: SIC in IBFD Systems }
As mentioned in Section \ref{ch2}-A, the SIC in IBFD system is a prime example application that demands high performance ICA architectures \cite{Ref_Fouda_2020,Ref_Lu_2020}. In this case, the ICA algorithm is applied to separate the self-interference signal and the signal of interest. An experimental platform in Matlab for the SIC in IBFD system is set up in \cite{Ref_Fouda_2020,Ref_Lu_2020}. This system emulates a WiFi data structure where a data frame contains 512 symbols and 48 subcarriers are employed. According to the normalized execution time reported in Table~\ref{Tab:Timing_ratio}, the preprocessing time should be less than 29.5\textmu s in this system. As a result, considering the latency of 78.33~\textmu{}s as shown in Table~\ref{Tab:Implementation}, three preprocessors would be needed for achieving real-time preprocessing. This further pushes the need for designing a more accelerated preprocessor architecture. Moreover, the proposed architecture is configurable to support different sample lengths and applies to different types of data frames in wireless communications.  
}
\section{Conclusion}
\label{ch6}

This paper presents a high-throughput and highly efficient configurable preprocessor for the ICA algorithm. Moreover, a high-performance MMA is also described. The proposed MMA architecture uses time-multiplexed processing, which greatly increases the efficiency of hardware utilization. Furthermore, the novel processing flow of the proposed preprocessor is highly optimized, so that the centering, the calculation of the covariance matrix, and the EVD are conducted in parallel and are pipelined. Thus, the processing throughput is maximized while the required number of hardware elements can be minimized. The proposed ICA preprocessor is designed and implemented with a circuit design flow. The performance estimates are based on post-layout evaluations. The proposed preprocessor achieves a throughput of 40.7~kMatrices per second with a complexity of 73.3~kGE. Compared with prior work, the proposed preprocessor achieves the highest processing throughput and best efficiency.

\nocite{*}
\bibliographystyle{ref/IEEEtran}

\begin{IEEEbiography}
{Hsi-Hung Lu} received the B.Sc. and  M.Sc. degree from National Taiwan University of Science and Technology (NTUST), Taipei, Taiwan in 2017 and 2021.
\end{IEEEbiography}
\begin{IEEEbiography}
[{\includegraphics[width=1in,height=1.25in,clip,keepaspectratio]{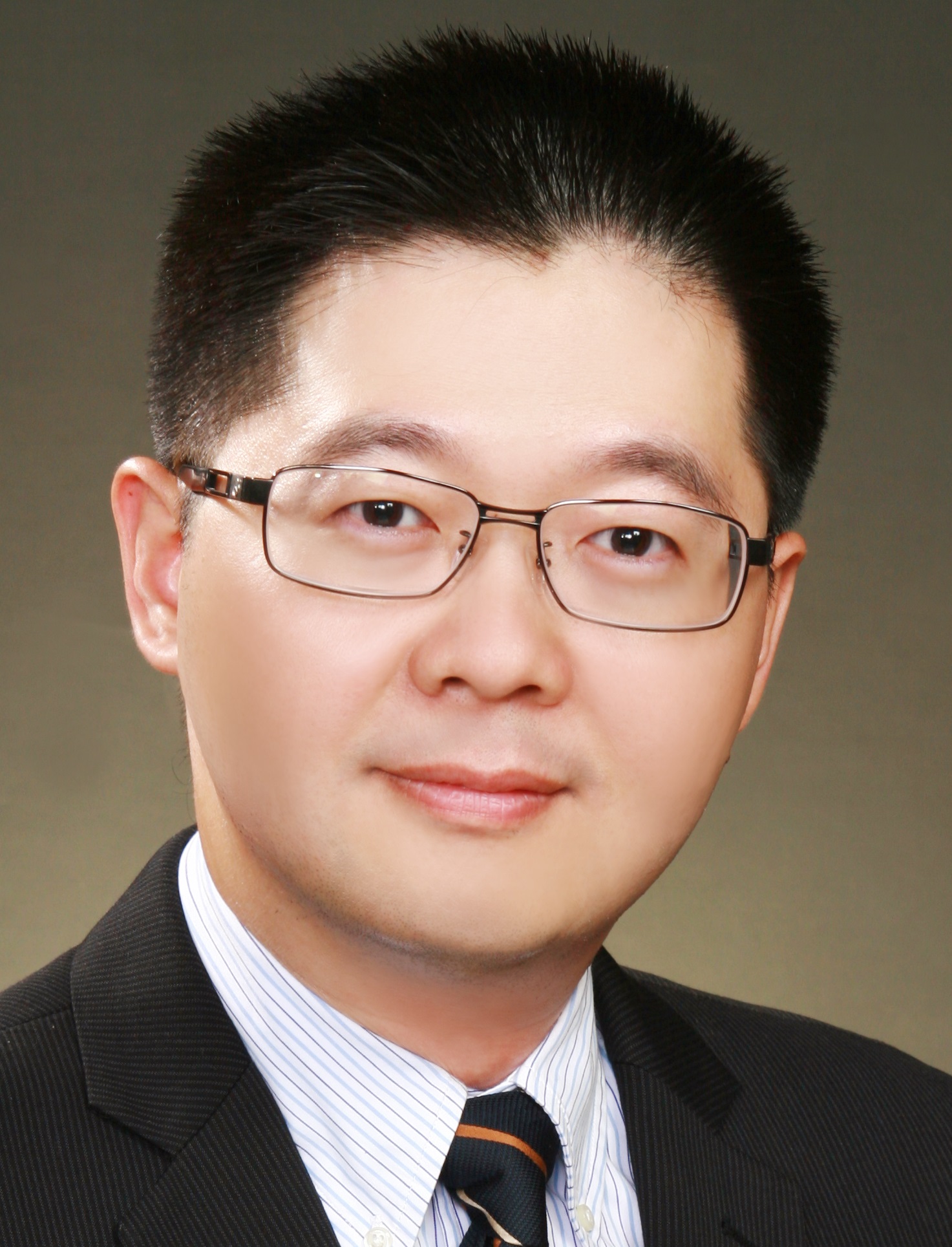}}]
{Chung-An Shen} received a B.Sc. degree from the National Taiwan University of Science and Technology (NTUST), Taipei, Taiwan, in 2000 and an M.Sc. degree from Ohio State University, Columbus, OH, USA, in 2003. He earned a Ph.D. degree from the University of California, Irvine, in 2012. Dr.~Shen joined the Department of Electronic and Computer Engineering at NTUST in 2012, where he is currently an associate professor. Dr.~Shen’s recent research activities are in low-power and highly efficient digital circuits and signal processing architectures for wireless communication systems. He collaborates with industry in the design and implementation of FPGA-based accelerators and offloading engines for 5G cellular networks.  Before joining NTUST, Dr.~Shen held several positions in industry and worked on digital signal processing and wireless system design.
\end{IEEEbiography}
\begin{IEEEbiography}
[{\includegraphics[width=1in,height=1.25in]{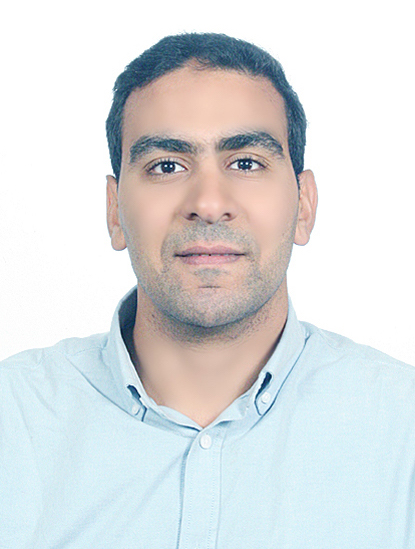}}]
{Mohammed E. Fouda} received a B.Sc. degree (Hons.) in Electronics and Communications Engineering and an M.Sc. degree in Engineering Mathematics from the Faculty of Engineering, Cairo University, Cairo, Egypt, in 2011 and 2014, respectively. Fouda received a Ph.D. degree from the University of California, Irvine, USA, in 2020. Currently, he works as an assistant researcher at the University of California, Irvine. Fouda has published more than 130 peer-reviewed journal and conference papers, one Springer book, and three book chapters. His H-index is 23, and he has been cited more than 1900 times. His research interests include analog AI hardware, neuromorphic circuits and systems, brain-inspired computing, memristive circuit theory, fractional circuits and systems, and analog circuits. He serves as a peer reviewer for many prestigious journals and conferences. He also serves as a review editor in \emph{Frontiers in Electronics} and the \emph{International Journal of Circuit Theory and Applications}, in addition to serving as a technical program committee member in many conferences. He was the recipient of the best paper award in ICM for 2013 and 2020 and the Broadcom foundation fellowship for 2016/2017.
\end{IEEEbiography}
\begin{IEEEbiography}
[{\includegraphics[width=1in,height=1.25in,clip,keepaspectratio]{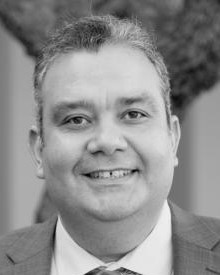}}]
{Ahmed E. Eltawil} (S’97–M’03–SM’14) is a Professor of Electrical and Computer Engineering at King Abdullah University of Science and Technology (KAUST) where he joined the Computer, Electrical and Mathematical Science and Engineering Division (CEMSE) in 2019. Prior to that he was with the Electrical Engineering and Computer Science Department at the University of California, Irvine (UCI) since 2005. At KAUST, he is the founder and director of the Communication and Computing Systems Laboratory (CCSL). His current research interests are in the general area of smart and connected systems with an emphasis on mobile systems. He received the Doctorate degree from the University of California, Los Angeles, in 2003 and the M.Sc. and B.Sc. degrees (with honors) from Cairo University, Giza, Egypt, in 1999 and 1997, respectively. Dr. Eltawil has been on the technical program committees and steering committees for numerous workshops, symposia, and conferences in the areas of low power computing and wireless communication system design. He received several awards, including the NSF CAREER grant supporting his research in low power computing and communication systems. He is a senior member of the IEEE and a senior member of the National Academy of Inventors, USA. In 2021, he was selected as “Innovator of the Year” by the Henry Samueli School of Engineering at the University of California, Irvine where he received United States Congress certificate of recognition for his contributions.
\end{IEEEbiography}
\end{document}